\documentclass[useAMS,usenatbib]{mn2e}
\usepackage{rotating}
\usepackage{journals}

\def\approxgt{\ifmmode \rlap{$>$}{}_{{}_{{}_{\textstyle\sim}}} \else%
$\rlap{$>$}{}_{{}_{{}_{\textstyle\sim}}}$\fi} 
\def\approxlt{\ifmmode \rlap{$<$}{}_{{}_{{}_{\textstyle\sim}}} \else%
$\rlap{$<$}{}_{{}_{{}_{\textstyle\sim}}}$\fi}

\LARGE \normalsize \title[Sidebands is 4U~1636--53]{Sidebands to the lower kilohertz QPO in
4U~1636--53}

\author[Jonker, M\'endez, \& van der Klis]  {P.G.~Jonker$^{1,2}$\thanks{email :
pjonker@cfa.harvard.edu}, M.~M\'endez$^{2,3}$, M.~van der
Klis$^4$\\ $^1$Harvard--Smithsonian Center for Astrophysics, 60 Garden Street,
Cambridge, MA~02138, Massachusetts, U.S.A.\\ $^2$SRON, National Institute for
Space Research, Sorbonnelaan 2, 3584~CA, Utrecht, The Netherlands\\
$^3$Astronomical Institute, Utrecht University, P.O.~Box 80000, 3508~TA,
Utrecht, The Netherlands\\ $^4$Astronomical Institute ``Anton Pannekoek'',
University of Amsterdam, Kruislaan 403, 1098 SJ Amsterdam, The Netherlands\\}

\begin{document}

\maketitle

\begin{abstract} \noindent In this Paper we report on further observations of
the third and fourth kilohertz quasi--periodic oscillations (QPOs) in
the power spectrum of the low--mass X--ray binary (LMXB)
4U~1636--53. These kilohertz QPOs are sidebands to the lower kilohertz
QPO. The upper sideband has a frequency 55.5$\pm$1.7~Hz larger than
that of the contemporaneously measured lower kilohertz QPO. Such a
sideband has now been measured at a significance $>6\sigma$ in the
power spectra of three neutron star LMXBs (4U~1636--53, 4U~1728--34,
and 4U~1608--52). We also confirm the presence of a sideband at a
frequency $\sim$55~Hz less than the frequency of the lower kilohertz
QPO. The lower sideband is detected at a 3.5$\sigma$ level, only when
the lower kilohertz QPO frequency is between 800 and 850 Hz. In that
frequency interval the sidebands are consistent with being symmetric
around the lower kHz QPO frequency. The upper limit to the rms
amplitude of the lower sideband is significant lower than that of the
upper sideband for lower kilohertz QPO frequencies $>$850
Hz. Symmetric sidebands are unique to 4U~1636--53. This might be
explained by the fact that lower kilohertz QPO frequencies as high as
800--850 Hz are rare for 4U~1728--34 and 4U~1608--52. Finally, we also
measured a low frequency QPO at a frequency of $\sim43$~Hz when the
lower kilohertz QPO frequency is between 700--850~Hz. A similar
low--frequency QPO is present in the power spectra of the other two
systems for which a sideband has been observed. We briefly discuss the
possibility that the sideband is caused by Lense--Thirring precession.

\end{abstract}


\section{Introduction} 

One of the important discoveries made with the Rossi X--ray Timing
Explorer (RXTE) satellite is that of the kilohertz quasi--periodic
oscillations (QPOs) in the power spectra of $\sim$20 accreting neutron
star low--mass X--ray binaries (see \citealt{va2000} for a
review). Kilohertz QPOs are caused by the motion of matter in regions
of spacetime within a few kilometres of the surface of accreting
neutron stars, where strong--field gravity is required to describe
such motion. They potentially allow one to detect strong--field
effects and to constrain the neutron star mass--radius
relation. Although there is as yet no agreement about the precise
physical mechanism underlying kilohertz QPOs, most models agree that
the frequency of one of the observed kilohertz QPOs reflects the
frequency of orbital motion at the inner edge of the accretion disc.

The basic phenomenology of kilohertz QPOs consists of two kilohertz
QPO peaks, which are separated by $\Delta\nu$=200--360\,Hz and which
move in frequency by up to 700\,Hz in general correlation with mass
accretion rate indicators (again see \citealt{va2000} for a
review). The highest observed frequency is $\sim$1330\,Hz
(\citealt{vafova2000}), corresponding in the case of a 1.4\,M$_\odot$
neutron star to an orbital radius as tight as 15\,km. Some of these
sources also show X--ray bursts, in which burst oscillations are seen
which last for a few seconds and have frequencies between 270 and
620\,Hz (see \citealt{2003strohbild} for a review). These oscillations
usually drift by 1-2\,Hz but are near the neutron star spin frequency
in each source (\citealt{2003Natur.424...42C}). In the millisecond
pulsar SAX~J1808.4--3658 $\Delta \nu$ has been found to be equal to
half the spin frequency, which disproves spin--orbit beat frequency
models (\citealt{2003Natur.424...44W}) and suggests instead that the
neutron star spin induces resonances in the disk flow, perhaps
involving the general relativistic epicyclic frequencies
(\citealt{2003PASJ...55..467A};
\citealt{2003Natur.424...44W}; \citealt{lambmiller2003}).

After extensive searches for additional kilohertz QPO peaks at theoretically
predicted frequencies which remained unsuccesful (e.g.~in Sco~X--1;
\citealt{2000MNRAS.318..938M}), \citet{2000ApJ...540L..29J} discovered 
sidebands at a frequency 50--65 Hz above the lower kilohertz QPO in
4U~1608--52, 4U~1728--34, and 4U~1636--53. Magnetospheric modulation
of the lower kilohertz QPO, a beat phenomenon taking place inside the
marginally stable orbit as well as Lense-Thirring precession are
possible explanations for these sidebands
(\citealt{2000ApJ...540L..29J}).
\citet{ps2000} demonstrated that the hydrodynamic disk mode model
(\citealt{psnor1999}) naturally produces a sideband at a frequency
near that observed plus several other, as yet unobserved peaks; no
other models have so far been able to explicitly accommodate the
sideband. The 50--65-Hz frequency difference between the lower
kilohertz QPO and the sideband frequency, the 'sideband separation'
$\Delta\nu_{SB}$, is different in each source and not identical to any
of the other QPO frequencies simultaneously observed in the 10--100 Hz
frequency range (\citealt{2000ApJ...540L..29J}).

In this Paper we find evidence for symmetric sidebands to the lower
kilohertz QPO in the atoll source 4U~1636--53.

\section{Observations, analysis, and results} 

We used all the proportional counter array (PCA) data from RXTE observations of
4U~1636--53 available to us at the beginning of 2004. Hence, we included the data used by 
\citet{2000ApJ...540L..29J}. The analysis we performed was the same as that done by
\citet{2000ApJ...540L..29J}. Below, a condensed description of the analysis steps is
given; more details can be found in \citet{2000ApJ...540L..29J}. 

Using 128~s--long segments of high time resolution PCA data (122$\mu$s
resolution), we calculated power spectra up to a Nyquist frequency of 4096~Hz
in an energy band of 2--60 keV. The power spectra were searched for lower
kilohertz QPOs which are narrow (full--width at half maximum, FWHM, less than
$\sim$10 Hz) and detected in 128~s at a significance larger than 2$\sigma$.
This resulted in a selection of 244~ksec of data. 

\begin{figure} \includegraphics[width=8cm]{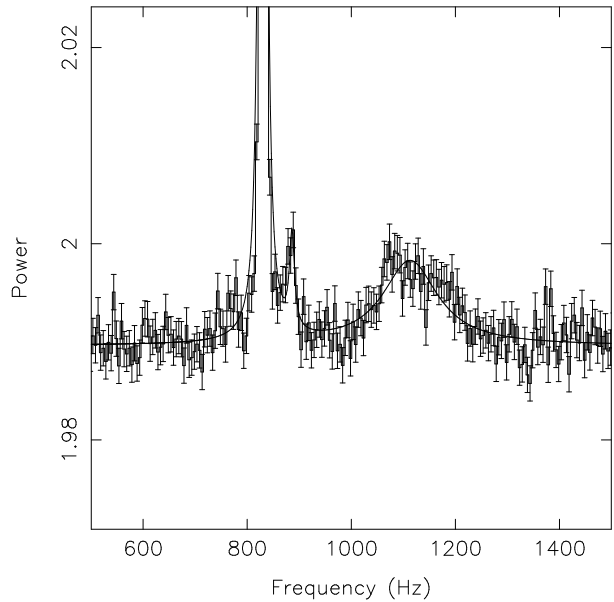} \caption{The
average power density spectrum of 4U~1636--53 showing the lower
kilohertz QPO (the peak is off--scale), the upper kilohertz QPO, and
the upper sideband to the lower kilohertz QPO. Due to the
shift--and--add method applied in the analysis only the frequency
differences are meaningfull (see text and references therein). The
drawn line represents the best fit of a function consisting of a
constant and three Lorentzians. The power in the Y--axis is given in
Leahy units.}
\label{sideb}
\end{figure}

The lower kilohertz QPO was traced using a dynamical power spectrum (e.g.~see plate 1 in
\citealt{bevava1996}) to visualise the time evolution of the QPO frequency. For each power
spectrum the lower kilohertz QPO peak was fitted in a range of 200~Hz centred on the
traced frequency using a function consisting of a constant plus a Lorentzian. This
provides us with a lower kilohertz QPO frequency measurement for each 128~s power
spectrum. We used the shift--and--add technique described by \citet{mevava1998b} to shift
each lower kilohertz QPO to a reference frequency. Next, the shifted, aligned, power
spectra were averaged. The average power spectrum was finally fitted in the range
512--2048 Hz so as to exclude the edges which are distorted due to the shifting. The fit
function consisted of a constant to fit the noise and three or four Lorentzians
representing the QPOs. The fitted average power spectrum is displayed in
Figure~\ref{sideb}.  Errors on the fit parameters were calculated using $\Delta
\chi^2=1.0$ (1$\sigma$ single parameter).

We clearly detected the two kilohertz QPOs and the sideband to the
lower kilohertz QPO. The sideband was detected at a significance of
6$\sigma$, confirming our previous $3\sigma$ detection of the sideband
in 4U~1636--53 (\citealt{2000ApJ...540L..29J}). The frequency
difference between the frequency of the lower kilohertz QPO and that
of the sideband, $\Delta\nu_{\rm SB}=55.5\pm1.7$~Hz.  The average
kilohertz QPO frequency separation, $\langle\Delta\nu=\nu_{\rm
upper}-\nu_{\rm lower}\rangle$, is 283$\pm$5~Hz. The FWHM of the
sideband, the lower kilohertz QPO and the upper kilohertz QPO is
12.7$\pm$2.7~Hz, 4.86$\pm$0.04~Hz, and 130$\pm$11~Hz,
respectively. Due to the fact that we have combined data with either
3, 4, or 5 active proportional counter units (PCUs) we cannot
determine the fractional rms amplitude of the QPO peaks from the
combined data. Therefore, we selected data where only 3, 4, or 5 PCUs
were active, we fitted those data sets separately. In those separate
fits we fixed the FWHM and the frequency of the QPOs to the values
found using the complete data set.  The fractional rms amplitude of
the sideband is consistent with what was found before (see
Table~\ref{side} and \citealt{2000ApJ...540L..29J}).

\begin{figure} \includegraphics[width=8cm]{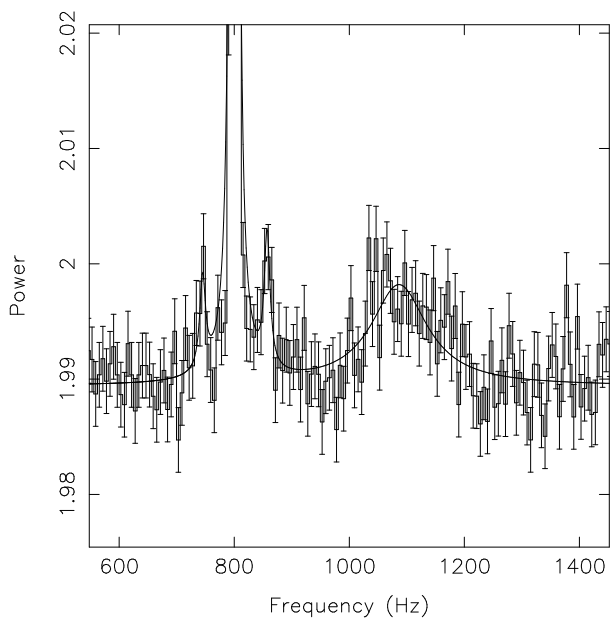} \caption{The
average power density spectrum of 4U~1636--53 selecting only lower kilohertz
QPO frequencies between 800-850~Hz. The two kilohertz QPOs and the sideband on
the high frequency side of the lower kilohertz QPO are visible. In addition, a
sideband on the low frequency side of the lower kilohertz QPO can be discerned
($3.5\sigma$ single trial). Due to the shift--and--add method applied in the
analysis only the frequency differences are usefull (see text and
references therein). The drawn line represents the best fit of a function
consisting of a constant and four Lorentzians.  } \label{twosideb} \end{figure}

In order to search for changes in the sideband separation frequency
$\Delta\nu_{\rm SB}$ we averaged shifted power spectra based on the unshifted
frequency of the lower kilohertz QPO in three frequency bands; a lower
kilohertz QPO frequency between 700--800 Hz, 800--850 Hz, and above 850 Hz.
Interestingly, the average power spectrum in the lower kilohertz QPO frequency
interval 800-850~Hz showed evidence for two sidebands located symmetrically
around the lower kilohertz QPO peak (see Figure~\ref{twosideb}). The results of
fits to these averaged power spectra are given in Table~\ref{side}.

\begin{table*}

\caption{Properties of the lower and upper kilohertz QPO ($\nu_1$ and $\nu_2$,
respectively), the sidebands to the lower kilohertz QPO, and the low frequency QPO. The
properties  displayed below the last horizontal line have been determined without using
the shift--and--add technique. The FWHM of the lower sideband has in all cases been fixed
to that of the upper sideband.}

\label{side}
\begin{center}
\begin{tabular}{lcccc}

       & All frequencies & 700--800 Hz & 800--850 Hz & 850--920 Hz \\ 
\hline
Amount of data  (ksec) & 244 & 64 &  80 & 99 \\ 
Mean $\nu_1\pm$stan. dev. (Hz) & 830$\pm$50 & 763$\pm$25 & 828$\pm15$ & 877$\pm17$ \\
$\Delta\nu$ (Hz)  & 283$\pm$5 & 325$\pm$6 & 286$\pm$9 & 249$\pm$4 \\
$\Delta\nu_{\rm SB, lower}$ (Hz) & -- & -- & 55.1$\pm$2.5 & -- \\
$\Delta\nu_{\rm SB, upper}$ (Hz) & 55.5$\pm$1.7 & 47.7$\pm$3.5 & 57.0$\pm2.1$ & 56$\pm$2 \\
rms amplitude $\nu_1$ (\%)& 7.54$\pm$0.01 & 7.94$\pm$0.04 & 7.86$\pm$0.02 & 7.12$\pm$0.03 \\ 
FWHM $\nu_1$ & 4.86$\pm$0.04 & 5.4$\pm$0.1 & 4.45$\pm$0.06 & 4.90$\pm$0.06\\
rms amplitude $\nu_2$(\%) & 4.03$\pm$0.11 & 4.5$\pm$0.3 & 4.0$\pm$0.2 & 3.4$\pm$0.2 \\
FWHM $\nu_2$ & 130$\pm$11 & 104$\pm$15 & 120$\pm$18 & 87$\pm$13\\
rms amplitude $\nu_{\rm SB, lower}$ (\%) & $<$0.9$^a$ & $<$1.5$^a$ & 1.2$\pm$0.2 & $<$0.8$^a$ \\
rms amplitude $\nu_{\rm SB, upper}$ (\%)& 1.3$\pm$0.1& 1.7$\pm$0.2 & 1.4$\pm$0.2 & 1.2$\pm$0.1\\
FWHM $\nu_{\rm SB, upper}$ & 12.7$\pm$2.7& 14$\pm$5& 12$\pm$4 & 10$\pm$4 \\
\hline
rms amplitude $\nu_{\rm low}$ (\%)& -- & 2.7$\pm$0.2 & 2.3$\pm$0.1 & $<$1.6$^a$ \\
FWHM $\nu_{\rm low}$ (Hz) & -- & 17$\pm$3 & 20$\pm$4 & -- \\
Frequency $\nu_{\rm low}$ (Hz) & -- &  42.1$\pm$0.8 & 43.5$\pm$1.6& -- \\

\end{tabular}
\end{center}
{\footnotesize$^a$ 95\% upper limits (determined using $\Delta \chi^2=2.71$).}\\
\end{table*}

\begin{figure} \includegraphics[width=8cm]{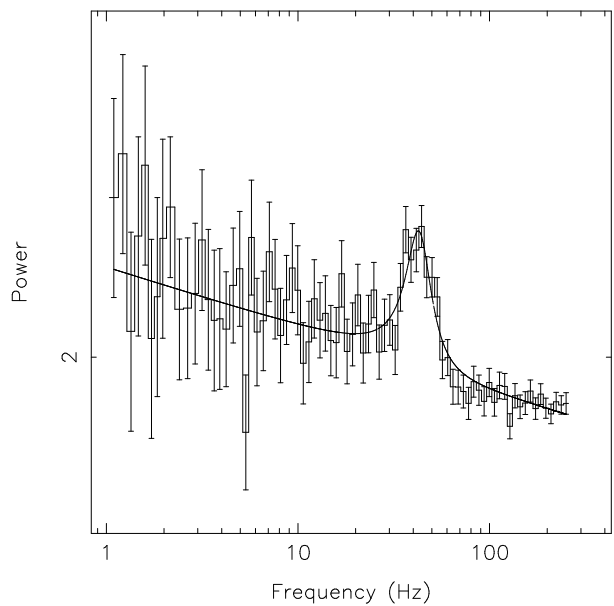} \caption{The
low frequency part of the average power density spectrum of
4U~1636--53 selecting only lower kilohertz QPO frequencies between
700-850~Hz. A low frequency QPO at a frequency of $\sim43$~Hz is
visible. The drawn line represents the best fit of a function
consisting of a power law to fit the underlying red noise continuum
and a Lorentzian to fit the low frequency QPO. We did not apply any
shift to produce this power spectrum.  } \label{lowfreq} \end{figure}

Adding more data with respect to the analysis performed by 
\citet{2000ApJ...540L..29J}, we now find a QPO at low frequencies 
($\sim$43 Hz, see Figure~\ref{lowfreq}). In order to compare the
properties of this low--frequency QPO with that of the kilohertz QPOs
and the kilohertz QPO separation frequencies we fitted the average
unshifted power spectrum in the frequency range from 1/16--256 Hz in
the three lower kilohertz QPO frequency intervals as shown in
Table~\ref{side}.  In the frequency range 850 Hz and above the low
frequency QPO was undetectable (see Table~\ref{side}).

\section{Discussion}

Following the discovery by \citet{2000ApJ...540L..29J} of a new,
third, kilohertz QPO in the power spectra of the three atoll--type
low--mass X--ray binaries 4U~1608--52, 4U~1728--34, and 4U~1636--53 we
obtained additional data with the RXTE satellite of 4U~1636--53 with
the goal to study in detail the properties of this third kilohertz
QPO, also known as the sideband to the lower kilohertz QPO. We
selected 4U~1636--53 since in that source there is evidence for the
presence of two sidebands located symmetrically around the lower
kilohertz QPO peak frequency.

We confirm the presence of a sideband at a frequency 55.5$\pm1.7$~Hz
higher than that of the lower kilohertz QPO frequency in 4U~1636--53
at a 6 $\sigma$ significance level. Furthermore, the addition of the
extra data allowed us to investigate the sideband(s) as a function of
the lower kilohertz QPO frequency. We found that when the lower
kilohertz QPO has a frequency in the range 800--850~Hz there is
significant evidence (3.5$\sigma$, single trial) for the presence of a
sideband at the lower frequency side of the lower kilohertz QPO. Selecting
lower kilohertz QPO frequencies $>$850 Hz and when combining all the
data, the rms amplitude of the lower sideband is significantly less
than that of the upper sideband. For the frequency range 700--800 Hz
the upper limit is consistent with the rms amplitude of the upper
sideband.

In the framework of the sonic--point and spin--resonance model of
\citet{lambmiller2003}, the sideband is unlikely to be caused by 
Lense--Thirring precession at the radius where the spin--resonance
occurs (near the orbital radius with a Kepler frequency of
$\nu_{spin}/2$), since for neutron star spin frequencies such as those
considered here, the Lense--Thirring precession frequency at that
radius is $\approx$2 Hz, this is much too low. It is possible though
that the radiation pattern responsible for the upper kilohertz QPO is
modulated by the Lense--Thirring precession at the sonic--point
radius, this modulation would then in turn modulate the formation of
the lower kilohertz QPO. In such a scenario the upper kilohertz QPO
should also display sidebands. The fact that these have not yet been
detected can be explained by the large FWHM of the upper kilohertz
QPO, which precludes the detection of such sidebands. The skewed shape
of the upper kilohertz QPO in the average power spectrum (see
e.g.~Fig~\ref{sideb}) can be explained by the fact that the upper
kilohertz QPO rms amplitude and $\Delta\nu$ change as a function of
kilohertz QPO frequency (e.g.~\citealt{mevawi1998}, Table~\ref{side}).

\begin{figure} \includegraphics[width=8cm]{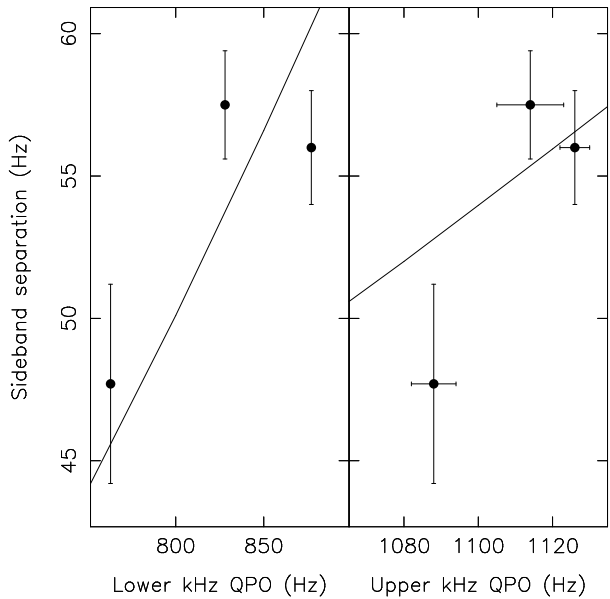} \caption{{\it Left
panel:} The sideband separation frequency plotted vs.~the lower
kilohertz QPO frequency, $\Delta\nu_{SB}$. The full line is the
quadratic relation between $\Delta\nu_{SB}$ and the lower kilohertz
QPO assuming the lower kilohertz QPO reflects the Keplerian frequency
and assuming the sideband is formed due to a modulation caused by
Lense-Thirring precession at that radius.  {\it Right panel:} same as
the left panel but for the upper kilohertz QPO.} \label{lense}
\end{figure}

If Lense--Thirring precession is responsible for the production of the
sideband, and if the sideband separation frequency $\Delta\nu_{SB}$,
reflects the Lense--Thirring precession frequency, $\Delta\nu_{SB}$
would be expected to change as the Keplerian frequency squared
($\Delta\nu_{SB}\propto\nu_{\rm K}^2$). In Figure~\ref{lense} we
plotted $\Delta\nu_{SB}$ as a function of the lower ({\it left panel})
and upper kilohertz QPO ({\it right panel}), respectively. The
sideband separation frequency $\Delta\nu_{SB}$ is consistent with
being constant when the lower/upper kilohertz QPO frequency changed by
$\sim100$/$\sim 38$~Hz, respectively. The best fitting quadratic
relation that would be appropriat assuming the lower/upper kilohertz
QPO reflects the Keplerian motion at a preferred radius in the disc
and $\Delta\nu_{SB}$ the Lense--Thirring precession frequency at that
same radius is also shown. The normalisation of the quadratic curve is
given by $\nu_{\rm LT}=\frac{8\pi^2 I \nu_{\rm K}^2 \nu_{s}}{c^2
M_{\rm NS}}$ ($\nu_{\rm LT}$ is the Lense--Thirring precession
frequency, I the moment of inertia of the neutron star, $\nu_{s}$ the
spin frequency of the star which is $\sim$581~Hz in case of
4U~1636--53 from \citealt{1997ApJ...486..355S}, $c$ the speed of
light, and $M_{\rm NS}$ the mass of the neutron star). The curve in
the left panel requires I$_{45}$/m to be 3.05$\pm$0.07, that in the
right panel 0.17$\pm$0.01, where I$_{45}$ and m are in units of
10$^{45}$ g cm$^{2}$ and M$_\odot$, respectively; the expected range
of this quantity is between 0.5 and 2 (\citealt{stvi1998}).

Recently, a model explaining the kilohertz QPOs has been proposed that
invokes the existence of a non--linear resonance between the vertical
and radial epicyclic frequencies in an accretion disc around a neutron
star (\citealt{2003PASJ...55..467A}). Some versions of this model also
include a resonance of these frequencies with the spin frequency of
the neutron star (\citealt{2004ApJ...603L..89K};
\citealt{2004ApJ...603L..93L}). The resonance model has been
introduced as a way to explain the existence of preferred values in
the distribution of QPO frequency ratios
(\citealt{2003PASJ...55..467A}; but see \citealt{belloni2004}) as well
as the commensurability of the neutron star spin and the kilohertz QPO
frequency separation. So far it has not been explored whether a
resonance mechanism could also explain the sidebands to the kilohertz
QPOs. For instance, it may be possible that the Lense--Thirring
precession at the radius at which the main resonance between the
vertical and radial epicyclic frequencies occurs produces a modulation
of the amplitude of the kilohertz QPOs. In principle, this modulation
should produce equally strong sidebands, and should also produce
sidebands on both QPOs. Alternatively, since for a rotating neutron
star the azimuthal (Keplerian) frequency is larger than the vertical
frequency, the sideband could be due to a resonance involving the
azimuthal frequency at a radius at which a resonance between the
vertical and radial epicyclic frequencies produce the upper and lower
kilohertz QPOs. For instance, in this scenario a resonance between the
azimuthal and radial frequencies would produce a QPO at a frequency
slightly higher than that of the upper kilohertz QPO. The advantage of
this mechanism compared to an amplitude modulation is that it could
in principle explain the fact that the upper sideband is stronger than
the lower sideband. However, it remains unclear how a QPO at a
frequency $\approx55$~Hz higher than that of the lower kilohertz QPO
would appear in this scenario.

\begin{figure} \includegraphics[width=7cm,angle=-90]{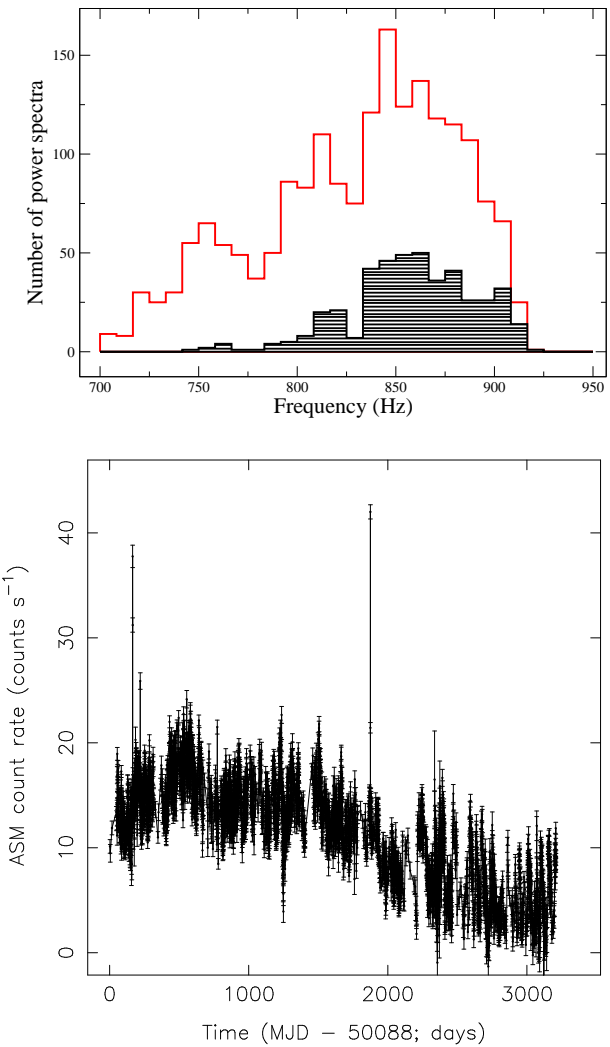}\quad
\includegraphics[width=8cm]{asm.1636.decay.ps} \caption{{\it Top panel:} The
distribution of the lower kilohertz QPO frequency for the data used by
Jonker et al.~(2000; black shaded area) and the data used in this work
(grey histogram). Note that the drawn grey distribution includes the
black shaded data. It is clear that, besides an increase in data
volume, a larger range in lower kilohertz QPO frequency was sampled
with the current data. The occurence of peaks in the frequency
distrubution can be explained by a random walk in kilohertz QPO
frequency (see for instance \citealt{belloni2004}). {\it Bottom
panel:} The ASM count rate history of the source 4U~1636--53. A clear
decay in overall brightness can be seen starting around MJD
$\sim$51590. The strong narrow peaks are caused by the occurence of
superbursts (\citealt{2001ApJ...554L..59W};
\citealt{2002ApJ...577..337S}). }
\label{distri} \end{figure}

Previously, 4U~1636--53 was the only source out of the three sources where a
sideband to the lower kilohertz QPO that showed no low--frequency QPO
(\citealt{2000ApJ...540L..29J}). Now, with the additional RXTE data, we could
identify a low--frequency QPO at a frequency of $\sim$43~Hz, similar to the
one found in 4U~1728--34 and 4U~1608--53 (see \citealt{2000ApJ...540L..29J}).
The probable reason why this low--frequency QPO was not observed before is
that the source did not sample the states where this QPO is present often.
Much of the new data were obtained in the state showing the low--frequency
QPO. From Table~\ref{side} it can be seen that the lower kilohertz QPO
frequency has to be below 850 Hz for the low--frequency QPO to be detected.
Comparing the observed frequency distributions of the data set used by
\citet{2000ApJ...540L..29J}  (Figure~\ref{distri} shaded area {\it top panel})
with that used in this work (Figure~\ref{distri} overall grey line {\it top
panel}) it is clear that besides an increase in the amount of data, the lower
kilohertz QPO frequencies in the range 700--850 Hz are much better sampled in
the new dataset. A likely explanation for the fact that the distribution of
the observed lower kilohertz QPO frequencies is different during the new RXTE
observations from that observed in the data set used by
\citet{2000ApJ...540L..29J}, is that the X--ray flux of 4U~1636--53 has been
decreasing over the last few years (see Fig.~\ref{distri} {\it bottom panel}),
allowing many of the additional new observations to be obtained at flux levels
and at lower kilohertz QPO frequencies which were sparsely observed before in
4U~1636--53 with RXTE

\section*{Acknowledgments}  

\noindent Support for this work was provided by NASA through Chandra
Postdoctoral Fellowship grant number PF3--40027 awarded by the Chandra X--ray
Center, which is operated by the Smithsonian Astrophysical Observatory for NASA
under contract NAS8--39073.

\end{document}